\documentclass[12pt,a4paper]{article}
\usepackage[in,plain]{fullpage}
\usepackage[centertags]{amsmath}
\usepackage{amssymb}
\usepackage{epsfig}
\renewcommand{\today}{July 1996}
\begin{document}
\begin{titlepage}
\null
\vspace{5mm}
\begin{flushright}
\begin{tabular}{l}
UWThPh-1996-42\\
DFTT 34/96\\
hep-ph/9607372\\
\today
\end{tabular}
\end{flushright}
\vspace{3mm}
\begin{center}
\Large
\textbf{Neutrino mass spectrum
from the results\\
of neutrino oscillation experiments}\\[5mm]
\normalsize
S.M. Bilenky\\
Joint Institute for Nuclear Research, Dubna, Russia, and\\
INFN, Sezione di Torino,
Via P. Giuria 1, I--10125 Torino, Italy\\[3mm]
C. Giunti\\
INFN, Sezione di Torino, and Dipartimento di Fisica Teorica,
Universit\`a di Torino,\\
Via P. Giuria 1, I--10125 Torino, Italy\\[3mm]
and\\[3mm]
W. Grimus\\
Institute for Theoretical Physics, University of Vienna,\\
Boltzmanngasse 5, A--1090 Vienna, Austria\\
\vspace{10mm}
\textbf{Abstract}\\[3mm]
\begin{minipage}{0.8\textwidth}
All the possible schemes of neutrino mixing
with four massive neutrinos
inspired by the existing experimental indications
in favor of neutrino mixing
are considered
in a model independent way.
Assuming that in short-baseline experiments only one
mass-squared difference is relevant,
it is shown that the scheme with a neutrino
mass hierarchy
is not compatible with
the experimental results.
Only two schemes
with two pairs of neutrinos with close masses
separated by a mass difference
of the order of 1 eV
are in agreement with the results
of all experiments.
One of these schemes
leads to possibly observable effects in
$^3$H and $(\beta\beta)_{0\nu}$
experiments.
\end{minipage}
\end{center}
\end{titlepage}

\section{Introduction}
\label{Introduction}

Neutrino masses and neutrino mixing are natural and
plausible phenomena of modern gauge theories
(see, for example, Ref.\cite{gauge}).
However,
for the time being,
the values of the
neutrino masses and mixing angles cannot
be predicted on theoretical grounds.
The determination of these quantities is the key
problem of today's experimental neutrino physics.

At present there are several indications
in favor of neutrino masses and
mixing. One of the most important
indications comes from solar
neutrino experiments
(Homestake \cite{Homestake},
Kamiokande \cite{Kamiokande},
GALLEX \cite{GALLEX}
and SAGE \cite{SAGE}).
As it is well-known, in all four
presently operating solar neutrino experiments the observed event
rates are significantly smaller than the values predicted by the
Standard Solar Model (SSM)
\cite{SSM}.
Moreover,
if the survival probability of solar $\nu_e$'s is
equal to one,
the data of \emph{different} experiments
cannot be explained
even if the total neutrino fluxes
are considered as free parameters
\cite{phenomenological}.
Assuming the validity of the SSM,
the experimental data can be
described by the MSW matter effect \cite{MSW} for
$\Delta{m}^{2} \sim 10^{-5}\, \mathrm{eV}^2 $
\cite{SOLMSW}
or by vacuum oscillations in the case
of $\Delta{m}^{2} \sim 10^{-10}\, \mathrm{eV}^2 $
\cite{SOLVAC}
($\Delta{m}^{2}$
is the neutrino mass-squared difference).

The second indication in favor
of neutrino mixing comes from the data of
the
Kamiokande
\cite{Kamiokande-atmospheric},
IMB
\cite{IMB}
and Soudan
\cite{Soudan}
atmospheric neutrino experiments.
The ratio
of muon-like to electron-like events measured in these experiments
is less than the expected ratio. The sub-GeV and multi-GeV data of
the Kamiokande collaboration can be explained by
$\nu_\mu \leftrightarrows \nu_\tau$ or $\nu_\mu \leftrightarrows \nu_e$
oscillations with $\Delta{m}^{2} \sim 10^{-2}\, \mathrm{eV}^2 $.

Finally, in the LSND
experiment \cite{LSND}
$ \bar\nu_e \, p \to e^+ \, n $
events produced by neutrinos
originating from $\pi^+$ and subsequent $\mu^+$ decays
at rest
were observed.
These events can be explained by
$ \bar\nu_\mu \leftrightarrows \bar\nu_e $
oscillations with
$ \Delta{m}^{2} \sim 1 \, \mathrm{eV}^2 $.

Therefore,
from the existing experimental data we have
three different indications in favor of non-zero
neutrino masses,
each with a particular scale of $\Delta{m}^{2}$.
However,
we must also take into account
the fact that
in several short-baseline experiments
neutrino oscillations were not observed.
The resuls of these experiments
allow to exclude
large regions in the space of the neutrino oscillation
parameters.

In the present paper we will examine
what information
on the neutrino mass spectrum
can be inferred from the results
of all short-baseline
neutrino oscillation experiments
if we also take into account
the results of the solar and atmospheric neutrino experiments.
We will present a general discussion
which does not assume
a definite model of neutrino mixing.
We will show that
the experimental results
favor two rather particular possibilities
for the neutrino mass spectrum.

We will start with the presentation of the
general formulas for
the neutrino transition probabilities in
short-baseline neutrino oscillation experiments
(for details see Ref.\cite{BGKP}).
Our basic
assumption is that
only one neutrino mass-squared difference is
relevant for
short-baseline neutrino oscillations.
This assumption means that the neutrino mass
spectrum consists of two groups of
close masses,
separated by a mass difference in the eV range.
Denoting the neutrinos of the two groups by
$ \nu_1, \ldots , \nu_r $
and
$ \nu_{r+1}, \ldots , \nu_n $,
respectively, with masses
$ m_1 \leq \ldots m_r $
and
$ m_{r+1} \leq \ldots \leq m_n $,
we can quantify our assumption
by
\begin{equation} \label{ass}
\frac{\Delta{m}^{2}_{i1} L}{2p}
\ll
1
\quad \mbox{for}
\quad
i \leq r
\quad
\mbox{and} \quad \frac{\Delta{m}^{2}_{ni} L}{2p} \ll 1
\quad \mbox{for} \quad i \geq r+1
\;,
\end{equation}
where
$\Delta{m}^{2}_{ij} \equiv m^2_i - m^2_j$,
$L$ is the
distance between
the neutrino source and detector
and $p$ is the neutrino momentum.
We would like to
emphasize that the inequalities
(\ref{ass}) are well satisfied
for short-baseline experiments with the mass squared
differences relevant for
the explanation of
the solar and atmospheric neutrino experiments.

Under the assumption of validity of
the inequalities (\ref{ass}),
the amplitude of the transition
$ \nu_{\alpha} \rightarrow \nu_{\beta} $
is given by
\begin{equation} \label{amp}
\mathcal{A}_{\nu_{\alpha}\rightarrow\nu_\beta}
\simeq
\mathrm{e}^{-iE_1t}
\left\{
\delta_{\alpha\beta}
+
\sum_{i \geq r+1}
U_{\beta i}
U_{\alpha i}^*
\left[
\exp\!\left(-i\frac{\Delta{m}^{2} L}{2p}\right) - 1
\right]
\right\}
\;.
\end{equation}
Here $\Delta{m}^{2} \equiv m^2_n - m^2_1$,
$U$ is the unitary $ n
\times n $ mixing matrix and $\nu_{\alpha}$,
$\nu_{\beta}$ are any active or sterile neutrinos.
(Note that the number
of sterile states is $n-3$.)
From Eq.(\ref{amp}),
for the probability of
the transition
$\nu_{\alpha}\rightarrow\nu_{\beta}$
($\alpha\neq\beta$)
we obtain
\begin{equation} \label{P12}
P_{\nu_{\alpha}\rightarrow\nu_{\beta}}
=
\frac{1}{2}
A_{\alpha;\beta}
\left( 1 - \cos \frac{\Delta{m}^{2} L}{2p} \right)
\;,
\end{equation}
where the oscillation amplitude
$A_{\alpha;\beta}$
is given by
\begin{equation} \label{A12}
A_{\alpha;\beta}  =
4 \left| \sum_{i \geq r+1} U_{\beta i} U_{\alpha i}^* \right|^2
=
A_{\beta;\alpha}
\;.
\end{equation}
The survival probability of $\nu_{\alpha}$ is
calculated as
\begin{equation} \label{P11}
P_{\nu_{\alpha}\rightarrow\nu_{\alpha}}
=
1 - \sum_{\beta\neq\alpha}
P_{\nu_{\alpha}\rightarrow\nu_\beta}
=
1 - \frac{1}{2}
B_{\alpha;\alpha}
\left(1 - \cos \frac{\Delta{m}^{2} L}{2p} \right)
\;,
\end{equation}
where
\begin{equation} \label{B11}
B_{\alpha;\alpha}
=
\sum_{\beta \neq \alpha}
A_{\alpha;\beta}
=
4 \sum_{i \geq r+1} |U_{\alpha i}|^2
\left( 1 - \sum_{i \geq r+1} |U_{\alpha i}|^2 \right)
\;.
\end{equation}

Using the unitarity of the mixing matrix,
from Eqs.(\ref{A12}) and (\ref{B11})
we have also
\begin{align}
&
A_{\alpha;\beta}
=
4 \left| \sum_{i \leq r}
U_{\beta i} U_{\alpha i}^* \right|^2
\;,
\label{A121}
\\
&
B_{\alpha;\alpha}
=
4 \sum_{i \leq r} |U_{\alpha i}|^2
\left( 1 - \sum_{i \leq r} |U_{\alpha i}|^2 \right)
\;.
\label{B111}
\end{align}

The expressions (\ref{P12})--(\ref{B111})
describe the transitions between all possible
neutrino states,
whether active or sterile.
Let us stress that in the scheme
under consideration the oscillations in
all channels are characterized
by the same oscillation length
$
L_{\mathrm{osc}}
=
4 \pi p / \Delta m^2
$.

\section{Schemes with three massive neutrinos}
\label{3nu}

Neutrino oscillations
in the scheme with
three massive neutrinos
and a mass hierarchy
were considered in many papers
\cite{three,BBGK,CF96YM96}.
In Ref.\cite{BBGK}
it was shown that the results of the LSND experiment
are compatible with the data of all the other experiments
on the search for neutrino oscillations
and the data of solar
neutrino experiments
only if the element $|U_{\mu3}|$ of the
neutrino mixing matrix is large (close to one)
and $|U_{e3}|$, $|U_{\tau3}|$ are small.

In Refs.\cite{CF96YM96}
the possibility was considered that the same
$\Delta{m}^{2} \simeq 0.3\, \mathrm{eV}^2 $
is relevant for the LSND effect and
for the atmospheric
neutrino anomaly.
This possibility seems to be rather marginal.
In fact, from the results of the
LSND \cite{LSND} and Bugey \cite{Bugey95}
experiments it follows that
the transition amplitude
$A_{\mu;e}$ is less than
$ 4 \times 10^{-2} $.
Therefore,
in order to explain the sub-GeV Kamiokande data
one needs a large amplitude
$
A_{\mu;\tau}
\simeq
B_{\mu;\mu}
$.
Taking into account the limits for
$B_{\mu;\mu}$
that follow from the
CDHS
$\nu_\mu$
disappearance experiment \cite{CDHS84},
one finds
$ \Delta{m}^{2} \lesssim 0.4 \, \mathrm{eV}^2 $.
On the
other hand,
the results of the LSND and Bugey experiments require
$\Delta{m}^{2} \gtrsim 0.3\, \mathrm{eV}^2 $.
Thus,
the results of all short-baseline oscillation
experiments are compatible
with the Kamiokande sub-GeV data only
if the value of $\Delta{m}^{2}$
lies in the very narrow interval
$ 0.3 \, \mathrm{eV}^2
\lesssim \Delta{m}^{2} \lesssim
0.4 \, \mathrm{eV}^2 $
and
$A_{\mu;e} \simeq 3 \times 10^{-2}$.
At such large values of $\Delta{m}^{2}$
the cosine in the expression (\ref{P11})
practically
disappears
for atmospheric neutrinos
because of the averaging over energy and distance.
This would mean
that there cannot be an zenithal dependence
of the double ratio
\begin{math}
R = (\mu/e)_{\mathrm{data}}/(\mu/e)_{\mathrm{MC}}
\end{math}
of atmospheric
muon and electron events
($(\mu/e)_{\mathrm{MC}}$
is the Monte-Carlo
calculated ratio of muon and electron events
without neutrino oscillations).
However,
some zenithal dependence
of this double ratio
is indicated by the
multi-GeV Kamiokande
data.

The existence of an zenithal dependence
of the double ratio $R$ will
be checked soon by the on-going Super-Kamiokande
experiment \cite{SK}.
Let us also mention that
$\nu_\mu \leftrightarrows \nu_\tau$
oscillations with large amplitude
and $\Delta{m}^{2} \simeq 0.3\, \mathrm{eV}^2 $
could be tested by CHORUS \cite{CHORUS},
NOMAD \cite{NOMAD}
and especially by the COSMOS \cite{COSMOS} experiment.

Up to now we have discussed
a scheme with three neutrinos and a mass
hierarchy.
The other possible scheme with three neutrinos
is a scheme where the
spectrum has the form
$ m_1 \ll m_2 \lesssim m_3 $
\cite{threeinv}.
With the same arguments as above
it is possible to show
\cite{BGKP}
that this scheme is also disfavored
by the experimental results.

\section{Four massive neutrinos with a mass hierarchy}
\label{4nu1}

We will now assume that the anomaly in the sub-GeV
and multi-GeV Kamiokande atmospheric neutrino data
is due
to neutrino oscillations,
for which a scale of
$ \Delta{m}^{2} \sim 10^{-2} \, \mathrm{eV^2} $
is required
\cite{Kamiokande-atmospheric}.
In order to
accomodate
also the solar neutrino data and the LSND data,
it is necessary to assume that
the flavor neutrino fields
are superpositions of (at least) four massive neutrino fields.
We will consider now all possible schemes with four neutrinos
and the dominance of one $\Delta{m}^{2}$
in short-baseline experiments.
We will start with the case of
a neutrino mass hierarchy,
$ m_1 \ll m_2 \ll m_3 \ll m_4 $,
with $\Delta{m}^{2}_{21}$ and $\Delta{m}^{2}_{32}$ relevant
for the suppression of the flux of solar neutrinos and for the
atmospheric neutrino anomaly, respectively.
This case
corresponds to $n=4$ and $r=3$
in the formulas (\ref{P12})--(\ref{B111})
and
the oscillation amplitudes
are given by
\begin{align}
&
A_{\alpha;\beta}
=
4 | U_{\beta 4} |^2 | U_{\alpha 4} |^2
\;,
\label{07}
\\
&
B_{\alpha;\alpha}
=
4 |U_{\alpha 4}|^2 \left( 1 - |U_{\alpha 4}|^2 \right)
\;.
\label{08}
\end{align}

We will consider the range
$
0.3 \, \mathrm{eV}^2
\leq \Delta{m}^{2} \leq
10^3 \, \mathrm{eV}^2
$,
which covers the sensitivity of all short-baseline
experiments.
At any fixed value of $\Delta{m}^{2}$,
from the
exclusion plots of the Bugey \cite{Bugey95},
CDHS \cite{CDHS84} and CCFR \cite{CCFR84}
disappearance experiments we have
\begin{equation} \label{B0}
B_{\alpha;\alpha} \leq B_{\alpha;\alpha}^{0}
\qquad
( \alpha = e, \mu )
\;.
\end{equation}
The values of
$ B_{e;e}^{0} $
and
$ B_{\mu;\mu}^{0} $
can be obtained from the corresponding exclusion curves.
From Eqs.(\ref{08}) and (\ref{B0}) we find
that the elements
$|U_{\alpha4}|^2$
must satisfy one of the two inequalities
\begin{equation}
|U_{\alpha4}|^2 \leq a^{0}_{\alpha}
\qquad \mbox{or} \qquad
|U_{\alpha4}|^2 \geq 1 - a^{0}_{\alpha}
\qquad
( \alpha = e, \mu )
\;,
\end{equation}
where
(see Ref.\cite{BBGK})
\begin{equation} \label{a0}
a^{0}_{\alpha} = \frac{1}{2}
\left(1-\sqrt{1-B_{\alpha;\alpha}^{0}}\,\right)
\;.
\end{equation}
In the range of $\Delta{m}^{2}$
considered here $a^{0}_e$ and $a^{0}_\mu$ are
small
($ a^{0}_e \lesssim 4 \times 10^{-2} $,
$ a^{0}_\mu \lesssim 10^{-1} $).
We will show now that solar neutrino data and the atmospheric
neutrino anomaly exclude large values of $|U_{e4}|^2$ and
$|U_{\mu4}|^2$.
In fact,
the average probability of solar
neutrinos to survive is given by
(see Refs.\cite{SS92,BGKP})
\begin{equation} \label{Psol}
P^{\odot}_{\nu_e\rightarrow\nu_e}
=
\sum_{i=3,4} |U_{ei}|^4
+
\left( 1 - \sum_{i=3,4} |U_{ei}|^2 \right)^2
P^{(1;2)}_{\nu_e\rightarrow\nu_e}
\;,
\end{equation}
where $P^{(1;2)}_{\nu_e\rightarrow\nu_e}$ is the $\nu_e$
survival probability
due to the mixing of $\nu_e$ with $\nu_1$ and $\nu_2$.
If
$ |U_{e4}|^2 \geq 1-a^{0}_e $,
from (\ref{Psol})
we have
$ P^{\odot}_{\nu_e\rightarrow\nu_e} \gtrsim 0.92 $
for all solar neutrino energies.
Such a
large lower bound is not compatible with the solar neutrino
data.

\begin{figure}[p]
\begin{center}
\mbox{\epsfig{file=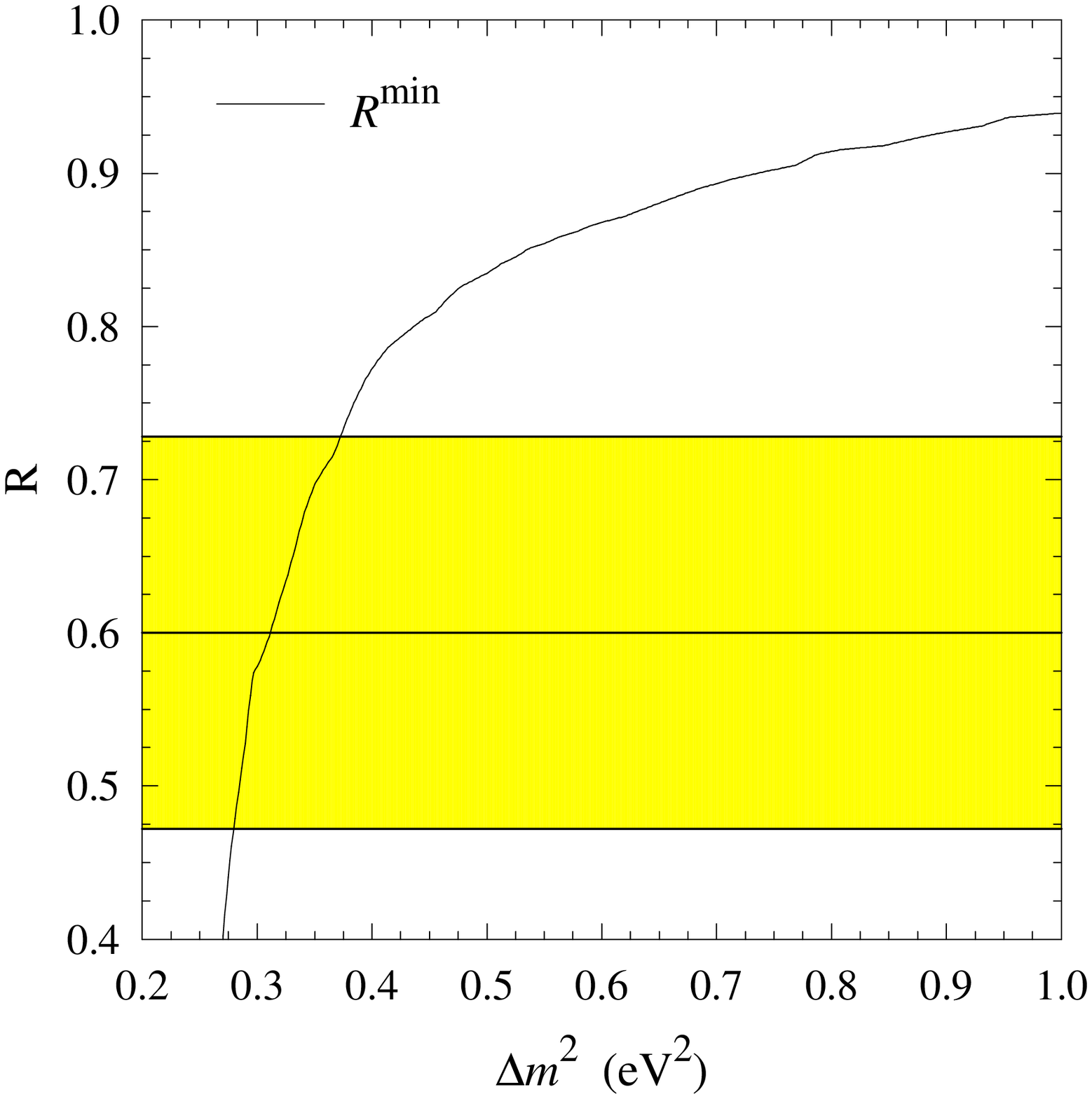,width=0.8\textwidth}}
\end{center}
\caption[Figure~\ref{fig1}]{\label{fig1}
Value of
$R^{\mathrm{min}}$
for the Kamiokande sub-GeV data
(see Eq.(\ref{002}))
as a function of
$\Delta{m}^2$.
The shadowed horizontal band
represents
the 90\% CL limits of the double ratio $R$
for the sub-GeV events
determined in the Kamiokande experiment
\cite{Kamiokande-atmospheric}.}
\end{figure}

The average probability of atmospheric $\nu_\mu$'s to
survive is given by
\begin{equation} \label{Patm}
P^{\mathrm{atm}}_{\nu_\mu\rightarrow\nu_\mu}
=
|U_{\mu4}|^4 +
\left( 1 - |U_{\mu4}|^2 \right)^2
P^{(1,2;3)}_{\nu_\mu\rightarrow\nu_\mu}
\;,
\end{equation}
where $P^{(1,2;3)}_{\nu_\mu\rightarrow\nu_\mu}$
is the $\nu_\mu$
survival probability\footnote{
$P^{(1,2;3)}_{\nu_\mu\rightarrow\nu_\mu}$
is given by
\begin{displaymath}
P^{(1,2;3)}_{\nu_\mu\rightarrow\nu_\mu}
=
1
-
2
\,
\frac
{\displaystyle
\left( |U_{{\mu}1}|^2 + |U_{{\mu}2}|^2 \right)
|U_{\mu3}|^2 }
{\displaystyle
\left( |U_{{\mu}1}|^2 + |U_{{\mu}2}|^2 + |U_{\mu3}|^2 \right)^2
}
\left(
1
-
\cos \frac{ \Delta{m}^2_{32} L }{ 2 p }
\right)
\;.
\end{displaymath}
}
due to the mixing of $\nu_\mu$ with
$\nu_3$ and $\nu_{2}$, $\nu_{1}$.
The double ratio
of atmospheric
muon and electron events
is given by
\begin{equation}
R
=
\frac
{
P_{\nu_\mu\to\nu_\mu}^{\mathrm{atm}}
+
r^{-1}
P_{\nu_e\to\nu_\mu}^{\mathrm{atm}}
}
{
P_{\nu_e\to\nu_e}^{\mathrm{atm}}
+
r
P_{\nu_\mu\to\nu_e}^{\mathrm{atm}}
}
\;,
\label{000}
\end{equation}
where $r$
is the ratio of muon and electron events
calculated
without neutrino oscillations.
For the Kamiokande sub-GeV events
$ r \simeq 1.57 $
and there is no zenithal dependence
of the double ratio $R$
\cite{Kamiokande-atmospheric}.
This means that the oscillatory terms in the
probabilities
$P_{\nu_\alpha\to\nu_\beta}^{\mathrm{atm}}$
($\alpha,\beta=e,\mu$)
in Eq.(\ref{000})
disappear because of the
averaging over energy and distance.
In this case
\begin{math}
P_{\nu_e\to\nu_\mu}^{\mathrm{atm}}
=
P_{\nu_\mu\to\nu_e}^{\mathrm{atm}}
\leq
1
-
P_{\nu_\mu\to\nu_\mu}^{\mathrm{atm}}
\end{math}.
From Eq.(\ref{000}),
we obtain for $R$ the lower bound
\begin{equation}
R^{\mathrm{min}}
=
P_{\nu_\mu\to\nu_\mu}^{\mathrm{min}}
\;.
\label{002}
\end{equation}
Let us consider now
the case
$ |U_{\mu4}|^2 \geq 1 - a^{0}_{\mu} $,
which implies
\begin{equation}
P^{\mathrm{min}}_{\nu_\mu\to\nu_\mu}
=
\left( 1 - a^{0}_{\mu} \right)^2
\;.
\label{003}
\end{equation}
The solid curve in Fig.\ref{fig1}
depicts the corresponding value of
$R^{\mathrm{min}}$
for the Kamiokande sub-GeV events
as a function of
$\Delta{m}^2$.
The shadowed horizontal band
in Fig.\ref{fig1} represents
the 90\% CL limits of the double ratio $R$
for the sub-GeV events
determined in the Kamiokande experiment
\cite{Kamiokande-atmospheric}.
From Fig.\ref{fig1}
one can see that,
for
$ \Delta{m}^2 \gtrsim 0.4 \, \mathrm{eV}^2 $
and
$ |U_{\mu4}|^2 \geq 1 - a^{0}_{\mu} $,
the value of
$R^{\mathrm{min}}$
is bigger than the experimental upper limit for $R$.

With the argument presented above
we cannot exclude large values of
$ |U_{\mu4}|^2 $
in the small interval
$ 0.3  \, \mathrm{eV}^2
\lesssim \Delta{m}^2 \lesssim
0.4 \, \mathrm{eV}^2 $.
However,
if the atmospheric neutrino anomaly
is mainly due to
a deficit of $\mu$-like events
(as indicated by a comparison of the experimental data
\cite{Kamiokande-atmospheric,IMB,Soudan}
with the existing calculations of the atmosperic neutrino
fluxes
\cite{Stanev},
with the exception of the
calculation presented in Ref.\cite{Naumov}),
large values of
$ |U_{\mu4}|^2 $
are incompatible with the data
for all values of
$\Delta{m}^2$.
This is connected with the fact
that for large values of
$ |U_{\mu4}|^2 $
the coefficient
$ \left( 1 - |U_{\mu4}|^2 \right)^2 $
of the survival probability
$P^{(1,2;3)}_{\nu_\mu\rightarrow\nu_\mu}$
in Eq.(\ref{Patm})
is very small
and it is not possible to explain
the zenithal dependence observed in the
Kamiokande multi-GeV data
\cite{Kamiokande-atmospheric}
(for example,
at
$ \Delta{m}^2 = 0.3 \, \mathrm{eV}^2 $
we have
\begin{math}
\left( 1 - |U_{\mu4}|^2 \right)^2
\lesssim 0.06
\end{math}).

\begin{figure}[p]
\begin{center}
\mbox{\epsfig{file=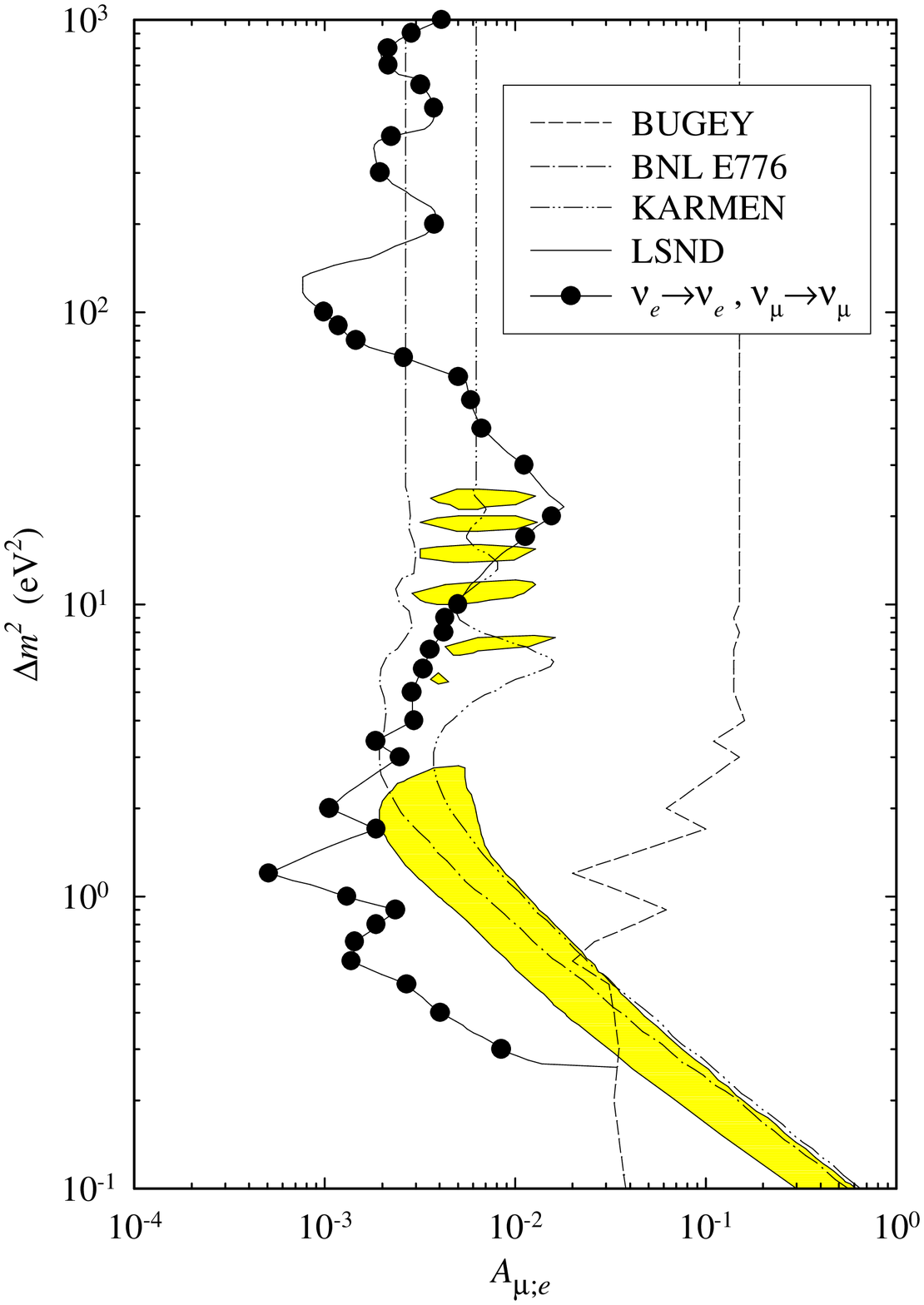,width=0.8\textwidth}}
\end{center}
\caption[Figure~\ref{fig2}]{\label{fig2}
Exclusion regions al 90\% CL in the
$ A_{\mu;e} $--$ \Delta m^2 $
plane
for small
$ \left| U_{e4} \right|^2 $
and
$ \left| U_{\mu4} \right|^2 $
in the model with mixing of four neutrinos
and a mass hierarchy
discussed in Section~\ref{4nu1}.
The regions excluded by
the BNL E776 and KARMEN
$ \nu_\mu \to \nu_e $
appearance experiments
are bounded by the dash-dotted and dash-dot-dotted curves,
respectively.
The dashed line represents
the results of the Bugey experiment.
The curve passing through the circles
is obtained
from the results
of the Bugey, CDHS and CCFR84 experiments
using Eq.(\ref{limit}).
The region allowed by the LSND experiment
is shown as
the shadowed region limited by the two solid curves.}
\end{figure}

Thus, using the results of the reactor and accelerator
disappearance
experiments and taking into account
the solar neutrino data and the
atmospheric neutrino anomaly,
we come to the conclusion that
\begin{equation} \label{bounds}
| U_{e4} |^2 \leq a^{0}_e
\qquad \mbox{and} \qquad
| U_{\mu4} |^2 \leq a^{0}_\mu
\;.
\end{equation}
We will consider now
$\nu_\mu \leftrightarrows \nu_e$ oscillations.
From
Eqs.(\ref{07}) and (\ref{bounds}) we have
\begin{equation} \label{limit}
A_{\mu;e}
=
4
| U_{e4} |^2
| U_{\mu4} |^2
\leq
4 a^{0}_e a^{0}_\mu
\;.
\end{equation}
Thus,
the upper bound for the amplitude
$A_{\mu;e}$
is quadratic in the small quantities
$a^{0}_e$, $a^{0}_\mu$,
and
$\nu_\mu \leftrightarrows \nu_e$ oscillations
must be strongly suppressed if there is a neutrino mass
hierarchy.
In Fig.\ref{fig2} the limit (\ref{limit})
is presented as the
curve passing through the circles.
The 90\% CL exclusion regions
found in the
$\bar\nu_e$ disappearance Bugey
experiment and in the $\nu_\mu \rightarrow \nu_e$
appearance BNL E776 \cite{BNLE776} and
KARMEN \cite{KARMEN} experiments
are limited in Fig.\ref{fig2}
by the dashed, dot-dashed and dot-dot-dashed curves,
respectively.
The shadowed
region in Fig.\ref{fig2} is
the region
of the parameters $\Delta{m}^{2}$ and
$A_{\mu;e}$ which is allowed at 90\% CL
by the LSND experiment.
It
is seen from Fig.\ref{fig2}
that the region allowed by LSND
is inside of  the
regions that are forbidden by the results of all the other
experiments.
Thus, we
come to the conclusion
that a mass hierarchy
of four neutrinos
is not compatible with the results
of all neutrino oscillation experiments.

\section{Four massive neutrinos with non-hierarchial mass spectra}
\label{4nu2}

If the neutrino masses satisfy the inequalities
\begin{equation} \label{spec}
m_1 \ll m_2 \lesssim m_3 \lesssim m_4
\;,
\end{equation}
with $\Delta{m}^{2}_{32}$ and $\Delta{m}^{2}_{43}$
relevant for the suppression of
the solar $\nu_e$'s and for the atmospheric neutrino anomaly,
respectively,
the short-baseline oscillation amplitudes are
given by Eqs.(\ref{07}) and (\ref{08})
with the change
$|U_{\alpha4}|^2 \rightarrow |U_{\alpha1}|^2$.
Arguments similar to those
presented in Section~\ref{4nu1} lead us to the
conclusion that the mass spectrum (\ref{spec})
is disfavored by
the experimental data.
In a similar manner one can
demonstrate that all possible schemes with mass spectra in which
three masses are clustered and one mass is separated from the
cluster by the
$ \sim 1 \, \mathrm{eV} $
gap needed for the
explanation of the LSND data are not compatible with the results
of all neutrino oscillation experiments.

Now we are left only with
two possible neutrino mass spectra
in which the four neutrino
masses appear in two pairs
separated by
$ \sim 1 \, \mathrm{eV} $:
\begin{equation} \label{AB}
\mbox{(A)}
\qquad
\underbrace{
\overbrace{m_1 < m_2}^{\mbox{atm}}
\ll
\overbrace{m_3 < m_4}^{\mbox{solar}}
}_{\mbox{LSND}}
\qquad \mbox{and} \qquad
\mbox{(B)}
\qquad
\underbrace{
\overbrace{m_1 < m_2}^{\mbox{solar}}
\ll
\overbrace{m_3 < m_4}^{\mbox{atm}}
}_{\mbox{LSND}}
\;.
\end{equation}
The possible effects
of these neutrino mass spectra have been
discussed in Refs.\cite{four,BGKP}.
We will show now that
the schemes with the spectra (A) and (B) are
compatible with the results of all
neutrino oscillation experiments.
Let us define the quantities
\begin{equation} \label{e17}
c_{\alpha} \equiv \sum_{i=1,2} |U_{\alpha i}|^2
\qquad (\alpha = e, \mu)
\;.
\end{equation}
For both schemes
(A) and (B)
the amplitude $B_{\alpha;\alpha}$ is given by
(see Eq.(\ref{B11})
with $n=4$ and $r=2$)
\begin{equation} \label{e18}
B_{\alpha;\alpha}
=
4 c_{\alpha} ( 1 - c_{\alpha} )
\;.
\end{equation}
From the results of reactor and accelerator
disappearance
experiments it follows that the parameters
$c_{\alpha}$ must
satisfy one of the two inequalities
\begin{equation} \label{e19}
c_{\alpha} \leq a^{0}_{\alpha}
\qquad \mbox{or} \qquad
c_{\alpha} \geq 1-a^{0}_{\alpha}
\qquad (\alpha = e, \mu)
\;,
\end{equation}
where $a^{0}_{\alpha}$ is given by Eq.(\ref{a0}).

Let us first consider the scheme (A).
For the probabilities of solar
$\nu_e$'s and atmospheric $\nu_\mu$'s to survive we have
\begin{equation} \label{sun}
P^{\odot}_{\nu_e\rightarrow\nu_e} = \sum_{i=1,2} |U_{ei}|^4 +
(1-c_e)^2 P^{(3;4)}_{\nu_e\rightarrow\nu_e}
\end{equation}
and
\begin{equation} \label{atm}
P^{\mathrm{atm}}_{\nu_\mu\rightarrow\nu_\mu} = (1-c_\mu)^2 +
c_\mu^2 P^{(1;2)}_{\nu_\mu\rightarrow\nu_\mu}
\;.
\end{equation}
If $ c_e \geq 1-a^{0}_e $,
from Eq.(\ref{sun}) it follows that
the survival
probability of solar $\nu_e$'s,
$P^{\odot}_{\nu_e\rightarrow\nu_e}$,
practically does not depend
on the neutrino energy
and
\begin{equation}
P^{\odot}_{\nu_e\rightarrow\nu_e} \gtrsim 0.5
\;.
\end{equation}
This is disfavored
by the solar neutrino data
\cite{KP96}.
If $ c_\mu \leq a^{0}_\mu $,
from Eq.(\ref{atm})
we have
\begin{equation}
P^{\mathrm{atm}}_{\nu_\mu\rightarrow\nu_\mu}
\geq
\left( 1 - a^{0}_{\mu} \right)^2
\;,
\end{equation}
which is not compatible with the Kamiokande atmospheric neutrino
data for the reasons discussed in Section~\ref{4nu1}.
Thus,
in order to accomodate the solar neutrino data and
the atmospheric neutrino anomaly,
from the four possibilities (\ref{e19})
we must choose
\begin{equation} \label{e24}
c_e \leq a^{0}_e
\qquad \mbox{and} \qquad
c_\mu \geq 1-a^{0}_\mu
\;.
\end{equation}

Let us consider now
$\nu_\mu\leftrightarrows\nu_e$
oscillations.
Using the Cauchy-Schwarz inequality,
from Eq.(\ref{A12})
(with $n=4$ and $r=2$)
and Eq.(\ref{e17}),
for both schemes (A) and (B)
we find
\begin{equation} \label{e25}
A_{\mu;e}
=
4
\left|
\sum_{i=1,2}
U_{ei} U_{{\mu}i}^{*}
\right|^2
\leq
4 c_e c_\mu
\;.
\end{equation}
From Eqs.(\ref{e24}) and (\ref{e25})
it follows that the upper bound for
$A_{\mu;e}$
is linear in the small quantity
$a^{0}_{e}$.
Since
$ a^{0}_{e} \gtrsim 5 \times 10^{-3} $
for all values of
$\Delta{m}^2$,
in the case of the scheme (A)
the limit (\ref{e25})
is compatible with the results
of the LSND experiment.

In the case of the scheme (B),
the solar neutrino problem
and the atmospheric neutrino anomaly
can be explained by neutrino oscillations
only if
\begin{equation} \label{e26}
c_e \geq 1 - a^{0}_e
\qquad \mbox{and} \qquad
c_\mu \leq a^{0}_\mu
\;.
\end{equation}
From Eqs.(\ref{e25}) and (\ref{e26})
it follows that the scheme (B)
is also compatible with the results
of the LSND experiment.

The schemes (A) and (B)
lead to different consequences
for the experiments on the
mesurement of the neutrino mass
through the investigation
of the end-point part of the $^3$H $\beta$-spectrum
and
for the experiments on the search for
neutrinoless double-$\beta$ decay
($(\beta\beta)_{0\nu}$).
In fact,
for the whole range of $\Delta{m}^2$
considered here
we have
\begin{alignat}{5} \label{e27}
&
\mbox{(A)}
& \qquad &
\sum_{i=3,4}
|U_{ei}|^2
\geq
1 - a^{0}_e
\;,
&
\\ \label{e28}
&
\mbox{(B)}
&&
\sum_{i=3,4}
|U_{ei}|^2
\leq
a^{0}_e
\;.
&
\end{alignat}
From Eq.(\ref{e27})
it follows that
in the case of the scheme (A)
the neutrino mass that enters in the
usual expression for the $\beta$ spectrum of
$^3$H decay
(see Ref.\cite{RK88})
is approximately equal to
the ``LSND mass'' $m_4$:
\begin{equation} \label{e29}
m_{\nu}(^3\mathrm{H})
\simeq
m_4
\;.
\end{equation}

If the scheme (B)
is realized in nature
and $m_1$, $m_2$ are very small,
the mass measured in
$^3$H experiments
is at least two order
of magnitude smaller than $m_4$.

If massive neutrinos
are Majorana particles,
$(\beta\beta)_{0\nu}$
decay is possible.
In the scheme (A),
the effective neutrino ``mass''
\begin{math} \displaystyle
\left| \left\langle m \right\rangle \right|
=
\left|
\sum_{i=1}^{n}
U_{ei}^2
m_{i}
\right|
\end{math}
that is measured in
$(\beta\beta)_{0\nu}$
decay
is given by
\begin{equation} \label{e30}
\left| \left\langle m \right\rangle \right|
\simeq
\left|
\sum_{i=3,4}
U_{ei}^2
\right|
m_4
\;.
\end{equation}
We have
\begin{equation}
\left| \left\langle m \right\rangle \right|
\simeq
m_{4}
\sqrt{
1
-
4
|U_{e4}|^2
\left( 1 - |U_{e4}|^2 \right)
\sin^2\phi
}
\;,
\label{e301}
\end{equation}
where
$\phi$
is the difference of the phases of
$U_{e3}$ and $U_{e4}$.
Depending on the value of
the phase $\phi$,
the quantity
$\left| \left\langle m \right\rangle \right|$
has a value in the range
\begin{equation} \label{e31}
\left| 2 |U_{e4}|^2 - 1 \right|
m_4
\lesssim
\left| \left\langle m \right\rangle \right|
\lesssim
m_4
\;.
\end{equation}
The upper and lower bounds
in Eq.(\ref{e31})
correspond,
respectively,
to the cases of equal and opposite
CP parities of $\nu_3$ and $\nu_4$
(see Ref.\cite{BGKP}).
From Eq.(\ref{e28})
it follows that,
if $m_1$ and $m_2$ are very small,
in the case of the mass spectrum (B)
the expected value of
$\left| \left\langle m \right\rangle \right|$
is very small.

Thus,
the experiments
on the investigation of the effects of neutrino masses
with the measurement of the end-point part
of the $\beta $-spectrum of $^3$H
and with the search for $(\beta\beta)_{0\nu}$ decay
could allow to
distinguish between the possibilities
(A) or (B)
for the neutrino mass spectrum.

\section{Conclusions}
\label{Conclusions}

In this paper we have discussed the possible form of the
neutrino mass spectrum
that can be inferred from
the results of all
neutrino oscillation experiments,
including the solar and atmospheric neutrino
experiments.
In this investigation we
assumed only
that one neutrino mass
squared difference is relevant in short-baseline
oscillation experiments.
We have argued that
it is unlikely
that all data can be fitted
with three neutrinos,
particularly if one includes the
Kamiokande multi-GeV results.

The experimental indications in favor of
neutrino mixing
coming from
the results of the solar and atmospheric neutrino
experiments and of the LSND experiment
imply that there are (at least)
three different scales of $\Delta{m}^2$'s,
about
$ 10^{-5} \, \mathrm{eV}^2 $,
$ 10^{-2} \, \mathrm{eV}^2 $ and
$ 1 \, \mathrm{eV}^2 $.
We have considered all the possible
schemes with four massive neutrinos
which provide these
three scales of $\Delta{m}^2$'s.

We have shown that
the results of the LSND experiment
are not compatible
with the limits obtained
by all the other neutrino oscillation experiments
in the case of a neutrino mass hierarchy
($ m_1 \ll m_2 \ll m_3 \ll m_4 $)
and
in the cases of neutrino mass spectra
in which three masses
are clustered in a group
and one mass is separated
from the cluster with a mass difference of
the order of 1 eV,
which corresponds to the range of
$\Delta{m}^2$
relevant for the oscillations
observed in the LSND experiment.

We have also shown
that only two possible spectra
of neutrino masses,
(A) and (B)
(see (\ref{AB})),
with two pairs of close
masses separated by a mass difference
of the order of 1 eV
are compatible with the results of
all neutrino oscillation experiments.
If the neutrino mass spectrum (A)
is realized in nature,
the neutrino mass that is measured
in $^3$H $\beta$-decay experiments
coincide with the
``LSND mass''.
If the massive neutrinos are
Majorana particles,
in the case of scheme (A),
the experiments on the search for
$(\beta\beta)_{0\nu}$ decay
have good chances to
obtain a positive result.

Finally,
we want to remark that,
if the experimental indications
in favor of
neutrino oscillations
are confirmed,
the neutrino mass spectrum is very different from
the mass spectra of quarks
and charged leptons.
This is, however, not so astonishing,
because at least four neutrinos,
one of which is sterile,
are necessary in order to
explain the results of all
neutrino oscillation experiments.

\begin{flushleft}
\Large \textbf{Note Added}
\end{flushleft}
After this work was finished
and reported at the Neutrino '96 Conference
in Helsinki,
the preprint TMUP-HEL-9605
by N. Okada and O Yasuda
appeared
(hepph/9606411).
Some of our results
are also contained in this paper.

\begin{flushleft}
\Large \textbf{Acknowledments}
\end{flushleft}
We would like to thank
J. Bernabeu,
G. Conforto,
F. Martelli,
L. Mikaelian
and
S. Petcov
for useful discussions.
S.M.B. would like to express his deep gratitude
to Prof. H. Pietschmann
for the kind hospitality at the
Institute for Theoretical Physics
of the University of Vienna.

\end{document}